\documentclass{emulateapj}
\shorttitle{{\sc Sunrise} overview}
\shortauthors{S.~K.~Solanki et al.}

\begin{document}

\title{{\sc Sunrise}: instrument, mission, data and first results}

\author{
S.~K.~Solanki\altaffilmark{1,8},
P.~Barthol\altaffilmark{1},
S.~Danilovic\altaffilmark{1},
A.~Feller\altaffilmark{1},
A.~Gandorfer\altaffilmark{1},
J.~Hirzberger\altaffilmark{1},
T.~L.~Riethm\"uller\altaffilmark{1},
M.~Sch\"ussler\altaffilmark{1},
J.~A.~Bonet\altaffilmark{2},
V.~Mart\'inez Pillet\altaffilmark{2},
J.~C.~del Toro Iniesta\altaffilmark{3},
V.~Domingo\altaffilmark{4},
J.~Palacios\altaffilmark{4},
M.~Kn\"olker\altaffilmark{5},
N.~Bello Gonz\'alez\altaffilmark{6},
T.~Berkefeld\altaffilmark{6},
M.~Franz\altaffilmark{6},
W.~Schmidt\altaffilmark{6},
and
A.~M.~Title\altaffilmark{7}
}

\altaffiltext{1}{Max-Planck-Institut f\"ur Sonnensystemforschung, Max-Planck-Str. 2, 37191 Katlenburg-Lindau, Germany.}
\altaffiltext{2}{Instituto de Astrof\'{\i}sica de Canarias, C/Via L\'actea s/n, 38200 La Laguna, Tenerife, Spain.}
\altaffiltext{3}{Instituto de Astrof\'{\i}sica de Andaluc\'{\i}a (CSIC),
Apdo. de Correos 3004, E-18080, Granada, Spain}
\altaffiltext{4}{Grupo de Astronom\'\i a y Ciencias del Espacio (Univ. de Valencia),
             E-46980, Paterna, Valencia, Spain}
\altaffiltext{5}{High Altitude Observatory, National Center for Atmospheric Research,
            P.O. Box 3000, Boulder CO 80307-3000, USA.\footnote{HAO/NCAR is sponsored by the NSF}}
\altaffiltext{6}{Kiepenheuer-Institut f\"ur Sonnenphysik, Sch\"oneckstr. 6, 79104 Freiburg, Germany.}
\altaffiltext{7}{Lockheed-Martin Solar and Astrophysical Lab., Palo Alto, USA}
\altaffiltext{8}{School of Space Research, Kyung Hee University, Yongin, Gyeonggi, 446-701, Korea}

\email{solanki@mps.mpg.de}

\begin{abstract}
The {\sc Sunrise} balloon-borne solar observatory consists of a 1m aperture
Gregory telescope, a UV filter imager, an imaging vector polarimeter, an
image stabilization system and further infrastructure. The first
science flight of {\sc Sunrise} yielded high-quality data that reveal
the structure, dynamics and evolution of solar convection, oscillations and
magnetic fields at a resolution of around 100 km in the quiet Sun. After a
brief description of instruments and data, first qualitative results are
presented. In contrast to earlier observations, we clearly see
granulation at 214 nm. Images in Ca~{\sc ii}~H display narrow, short-lived dark
intergranular lanes between the bright edges of granules. The very
small-scale, mixed-polarity internetwork fields are found to be highly
dynamic. A significant increase in detectable magnetic flux is found after
phase-diversity-related reconstruction of polarization maps, indicating that
the polarities
are mixed right down to the spatial resolution limit, and probably beyond.

\end{abstract}


\section{Introduction}

In order to understand the processes that govern
solar activity, we must disentangle how the magnetic field
interacts with the solar plasma and guides the conversion of energy between
its mechanical, magnetic, radiative, and thermal forms.

The photosphere represents the key interaction region:
thermal, kinetic and magnetic energy all are of the same order of magnitude
and transform most easily from one form into another. This interaction in
turn leads to the creation of a rich
variety of magnetic structures, from sunspots down to intense magnetic field
concentrations on a length scale of 100 km or less.
This structuring provides the need to obtain data with a homogeneous and
constant resolution of such length scales. One possibility to fulfil this
requirement is to fly a telescope carried by a stratospheric balloon.

Balloon missions aiming at high-resolution solar studies have a long history.
An early highlight was the 12-inch Stratoscope, which flew multiple times in 1957
(Schwarzschild 1959) and 1959 (Danielson 1961).
It produced high resolution images of solar granulation, the solar limb and of
sunspot fine structure, which for many years represented the state of the art.
A similar-sized instrument equipped with a spectrometer, the Spectrostratoscope,
flew in 1975, but mainly obtained imaging data of solar granulation
(Mehltretter 1976, 1978, Wittmann \& Mehltretter, 1977).
The Soviet Stratospheric Solar Station had multiple flights (e.g., Krat et al.
1970). The 50~cm telescope provided nearly diffraction limited
broad-band images of solar granulation and a sunspot (Krat et al. 1972). These
data led Karpinsky (1989) to propose large-scale restructuring of the solar
granulation on a short time scale.
The 80~cm diameter Flare Genesis telescope, equipped with a sophisticated
Fabry-P\'erot based vector magnetograph (Rust et al. 1996), flew in 1996 and
in 2000, achieving a spatial resolution of roughly
0.5~arcsec (Georgoulis et al. 2002) during the second flight. This experiment
resulted in the discovery of bipolar magnetic features moving towards a very
young sunspot, which were interpreted as small U-loops (Bernasconi et al. 2002;
Georgoulis et al. 2002).

In addition to achieving high resolution, stratospheric balloon flights also
allow the Sun to be explored in near ultraviolet radiation that
is strongly attenuated by the Earth's atmosphere. In 1970
and 1971, a 20~cm telescope flew on
a stratospheric balloon and recorded images between 200~nm and 460~nm
(Hers\'e\ 1979). Later, the Rasolba balloon experiment, a 30~cm telescope with an
ultraviolet spectrograph, obtained high resolution spectra in the wavelength
range between 190~nm and 295~nm, which includes the Mg~{\sc ii}~h and k lines (Samain
and Lemaire 1985, Staath and Lemaire 1995).

{\sc Sunrise} extends this tradition of telescopes carried by a
stratospheric balloon in order to study the Sun. It combines high spatial resolution
with sensitivity to ultraviolet radiation. At 1~m diameter, it is the largest solar
telescope so far to leave the ground. It is equipped with sophisticated post-focus
instruments, including a UV imager and a filter-based vector magnetograph.


%

\section{Instrumentation and Mission}

\subsection{Instrument}

The {\sc Sunrise} stratospheric balloon-borne observatory is composed of a
telescope, two post-focus science instruments (SuFI and
IMaX, see below), an Image Stabilization and Light Distribution (ISLiD) unit,
 and a Correlating Wave-front Sensor (CWS), carried in a gondola, which
 possesses pointing capability.

\subsubsection{Telescope}

The telescope is a Gregory-type reflector with 1~m clear aperture and an
effective focal length of close to 25~m.
A heat-rejection wedge at the prime focus reflects 99\%\ of the light from
the solar disk off to the side, reducing the heat load on the post-focus
 instruments to approximately 10~W.
 The secondary mirror is actively controlled in three degrees of freedom to
 compensate for thermo\-elastic deformations of the telescope during flight.
 The post-focus instrumentation rests on top of the telescope.
More details are given by Barthol et al. (2010).

\subsubsection{SuFI}

The {\sc Sunrise} Filter Imager (SuFI) provides images at violet and near
ultraviolet wavelengths. The wavelengths sampled
by SuFI are: 214~nm at 10~nm bandwidth, 300~nm at 5~nm bandwidth, 312~nm at 1.2~nm
bandwidth, 388~nm at 0.8~nm bandwidth and 396.8~nm (core of Ca~{\sc ii}~H) at 0.18~nm
bandwidth. A $2048\times 2048$ UV-enhanced CCD is employed, with a plate scale of
0.02 arcsec per pixel, on average (the plate scale is slightly wavelength
dependent).
In order to overcome aberrations due to thermoelastic deformations of the
telescope and any remaining seeing, a phase-diversity technique
(cf. Paxman et al. 1992; L\"{o}fdahl and Scharmer 1994) is used:
a special optical arrangement in front of the
detector delivers a nominally focused image on one half of the detector, while
the other half receives an image with a defocus of one wave at 214~nm.
Hence the field of view is $15\times 40$ arcsec$^2$. Both halves of the CCD
together provide sufficient information for post-facto removal of low-order
aberrations from the image.

A cadence of better than an image every 2~s can be achieved, depending on the
exposure time, to either take rapid time series at a given wavelength or to
switch wavelengths. Thus the 4 longer wavelengths can be cycled through
within 8~s.
Since the exposure time of the 214~nm was typically 30~s even at local noon,
owing to residual ozone at and above float altitudes,
the cadence was correspondingly lower whenever this wavelength was included.

A description of SuFI can by found in Gandorfer et al. (2010).

\subsubsection{IMaX}

The Imaging Magnetograph eXperiment (IMaX) operates
in the Fe~{\sc i} 525.02~nm line (a Zeeman triplet with Land\'e factor $g = 3$). Images
in polarized light covering $50\times 50$~arcsec$^2$ are recorded
at a spectral resolution of 85~m\AA, normally at 4 wavelengths within the spectral
line and 1 in the nearby continuum. The full Stokes vector in these 5
wavelengths at a noise level of $10^{-3}$ is obtained in 30 sec,  which is the
typical cadence for most of the observations. The number of wavelength points
(between 3 and 12) and of polarization states can be varied
to obtain a higher cadence, or a better rendering of the line profile shape.

The spectral resolution and sampling is achieved by using a thermally stabilized
tunable solid-state Fabry-P\'erot etalon in double pass together with a
narrowband, prefilter with a full width at half maximum of 0.1~nm.

Polarization states are isolated with the help of two nematic liquid-crystal
modulators operated at a frequency of 4~Hz,
which are switched between four states for full Stokes vector polarimetry.
A dual-beam approach is taken, with 2 synchronized 1k$\times 1$k CCD cameras.
After every observing run, a plate is inserted into the light path in front of
one of the cameras in order to obtain phase-diversity information for
post-facto reconstruction.
Detailed information on IMaX is provided by Mart\'{\i}nez Pillet et al. (2010).

\subsubsection{ISLiD}

The Image Stabilization and Light Distribution (ISLiD) unit allows simultaneous
observations with the two science instruments by distributing the radiation
according to wavelength (200--400~nm to SuFI; 525~nm to IMaX; 500~nm to CWS),
while preserving diffraction-limited performance as well as polarization information.
ISLiD contains a rapid, piezo-driven tip-tilt mirror, which stabilizes the
image on the science instrument foci
by damping all vibrations and motions acting at frequencies up to 60~Hz.
This mirror is controlled by CWS, which is optically stimulated by ISLiD.
ISLiD is described by Gandorfer et al. (2010).

\subsubsection{CWS}

The Correlating Wavefront Sensor (CWS) is a Shack-Hartmann type wavefront sensor with a lenslet array in a pupil image
that feeds a field-of-view of $12\times 12$ arcsec$^2$ on a high-speed camera.
It is used in two ways, as a fast correlation tracker to derive control
signals for a tip-tilt mirror, and as a slow wavefront sensor (defocus and coma) for active
alignment control of the telescope secondary mirror.
The CWS is described by Berkefeld et al. (2010).

\subsubsection{Gondola}

The {\sc Sunrise} gondola provides the housing for the telescope, instruments,
power supply etc. In addition, it is responsible for the precision pointing
of the telescope towards the Sun.
Situated approximately 100~m below the drifting balloon, the gondola
attitude control system nominally keeps the telescope orientation fixed to the Sun
within a range of less than $\pm 45$~arcsec. Within this range, ISLiD and
CWS are able to compensate residual motions and allow continuous observations.
Azimuthal control of the gondola is performed via a momentum transfer
unit at its top.

Electrical power is provided by large photovoltaic arrays housed in two
panels placed left and right of the telescope.
On the rear side of the gondola, the instrument control electronics are
mounted on two racks.

The two data storage containers collecting the science data are mounted
well-secured inside one of
the upper side trusses of the core framework, which provides protection and
easy access for data recovery after landing.
The complete payload has dimensions of 5.5~m in width and length, is about
6.4~m high and has a mass of 1919.6 kg (without CSBF equipment, ballast
etc.). More information on the gondola is provided by Barthol et al. (2010).

\subsection{Mission}

{\sc Sunrise} was flown on a zero-pressure stratospheric long-duration balloon,
launched and operated by the Columbia Scientific Ballooning Facility (CSBF).
It was launched on June 8, 2009 at 6:27 UT (08:27h local time) from ESRANGE
(67.89°N, 21.10°E) near Kiruna in northern Sweden.
It then floated westwards at a mean cruise altitude of 36~km and landed on
Somerset island (northern Canada), suspended on a parachute,
on June 13, 2009 at 23:47 UT. The payload remained in direct sunlight during the
entire flight. Since at float altitudes the payload was
above 99\%\ of the Earth's atmosphere, virtually seeing-free observations were
possible all the time. Also, the balloon stayed above most of the ozone in the
Earth's atmosphere, allowing high-resolution imaging in the UV at 214~nm,
300~nm and 312~nm.

The loss of high-speed telemetry relatively soon after reaching float altitude
(due to the failure of a rented commercial telemetry system), meant that no
full images could be downloaded during the entire mission, so that instrument
commissioning and operations had to be carried out practically blindly.


\section{Overview of the data recorded during the 2009 flight of {\sc Sunrise}}

The total observation time was 130 hours, in which IMaX acquired 415~GB of data
(480332 images) and SuFI recorded 790~GB (150288 images). 55685 of the SuFI images
were acquired while the CWS control loop was closed. 16128 of these closed-loop
images were collected near or at the solar limb ($\mu$~$<$~0.5).

During 23\% of the total time at float altitude the CWS loop was closed. We obtained
continuous time series longer than 1 min during 22\% and time series longer
than 10 min during 10\% of the total observation time. The longest time
series of SuFI data is 34 min, but only 19 min for the SuFI mode that includes
the shortest wavelength
of 214~nm. This wavelength is by far the most sensitive to the airmass along
the line-of-sight (LOS) and was only observed around local noon. The longest
IMaX time series lasts 32~min.

During data reduction, different versions of SuFI Data are generated.
Starting from the level-0 raw
data, level-1 data are produced that are fully reduced, but not phase-diversity
reconstructed. A first phase-diversity reconstruction using individual
wavefronts for each image results in level-2 data. Finally, level-3 data are
phase-diversity
reconstructed employing an averaged wavefront (see Hirzberger et al. 2010b for
more details on {\sc Sunrise}/SuFI data reduction). A brief summary of
the recorded SuFI data is given in Tables 1 and 2.

%

   \begin{table}
   \caption{SuFI observing modes.}
   \label{SufiModes}                                     
   \centering                                            
   \begin{tabular}{l l}                                  
   \hline                                                
   \noalign{\smallskip}
   SuFI mode                   & Fraction of \\
                               & observing time \\
   \hline
   \noalign{\smallskip}
   {$5\lambda$: 214, 300, 312, 388, 397~nm}  & {~8\%}    \\
   \noalign{\smallskip}
   {$4\lambda$: 300, 312, 388, 397~nm} & {56\%}    \\
   \noalign{\smallskip}
   {$3\lambda$: 300, 388, 397~nm}  & {~9\%}    \\
   \noalign{\smallskip}
   {$2\lambda$: mostly 388, 397~nm}     & {12\%}    \\
   \noalign{\smallskip}
   {$1\lambda$: mostly 397~nm}    & {15\%}    \\
   \noalign{\smallskip}
   \hline                                                
   \end{tabular}
   \end{table}

   \begin{table}
   \caption{SuFI Exposure times and numbers of images}
   \label{ExpTimeRanges}                                 
   \centering                                            
   \begin{tabular}{l r r r r}                                
   \hline                                                
   \noalign{\smallskip}
   Central     & Exp. time       & Exp. time     & No. images      & No. images \\
    $\lambda$  & $\mu$~$\ge$~0.5 & $\mu$~$<$~0.5 & $\mu$~$\ge$~0.5 & $\mu$~$<$~0.5 \\
   \noalign{\smallskip}
   [nm]        & [ms]            & [ms]          \\
   \hline
   \noalign{\smallskip}
   {214}    & {30000}      & {30000}    & {~~409}      & {~~~33}    \\
   {300}    & {100--500}   & {200--800} & {~8949}      & {~2649}    \\
   {312}    & {100--500}   & {200--800} & {~7372}      & {~2662}   \\
   {388}    & {~60--150}   & {~65--210} & {11560}      & {~2992}    \\
   {396.8}  & {250--1200}   & {750--1500} & {11267}      & {~7792}    \\
   \noalign{\smallskip}
   {$\Sigma$} & & & {39557}      & {16128}    \\
   \noalign{\smallskip}
   \hline                                                
   \end{tabular}
   \end{table}

IMaX observed in various modes that differ in the number of Stokes
parameters observed, the number of wavelength points and the number of
integrations at each wavelength. They are designated by the following
nomenclature: Letter L or V (L=longitudinal, i.e. only Stokes $I$ and $V$ are
observed, V=vector, i.e. full Stokes vector observed) followed by the number
of wavelength points (3, 5  and 12 were used on this flight) and the number
of integrations per wavelength. The most widely
used ("standard") mode is V5-6, i.e., Stokes vector measured in 5 wavelengths
(4 in the line and 1 in the continuum), with 6 images accumulated per wavelength
point. The cadence of the observations, the achieved noise level and the
amount of time for which each mode was observed is given in Table 3. Note
that the noise level refers to the Stokes $I$ continuum and is given for the
unreconstructed data. After reconstruction, the noise level increases by
roughly a factor of 3. The effective spatial resolution of the reconstructed
data at 0.15--$0.18$ arcsec, however, is considerably higher and results in
the resolution of both, small-scale magnetic and convective features
(Lagg et al. 2010, Khomenko et al. 2010).

The reduction of IMaX data and the various calibrations of the instrument
are described by Mart\'\i{nez} Pillet et al. (2010). Different versions of
IMaX data are produced. These include level-0 raw data, level-1 fully
reduced, but not reconstructed data and level-2 data, which were reconstructed
by deconvolution using a modified Wiener filter and the point spread function
(PSF) of the
optical system derived from calibration applying phase diversity.
The  reconstruction of the IMaX data always makes use of an averaged wavefront, so
that IMaX level-2 data correspond to SuFI level-3 data.

\begin{table}
   \caption{Time spans and cadences of various IMaX modes}
   \label{IMAXTable}                                 
   \centering                                            
   \begin{tabular}{l r r r }                                
   \hline                                                
   \noalign{\smallskip}
  Date & cadence & noise   &  Total \\
  Mode  &  [s] & [$I_c$] &  [min] \\
  \noalign{\smallskip}
  \hline                                                
   \noalign{\smallskip}
V5-6   &  33 & $0.0011$    &  553 \\
V5-3   &  18 &  $0.0015$   & 180 \\
V3-6   &  20 &  $0.0011$   &  13 \\
L12-2  &  31 &  $0.0013$   &  64 \\
L3-2   &  8 &  $0.0013$  &  42 \\
   \noalign{\smallskip}
   \hline                                                
   \end{tabular}
   \end{table}


\section{First results}

The Sun was extremely quiet during the entire flight of {\sc Sunrise}, so that
almost all of the {\sc Sunrise} data correspond to internetwork regions with
occasional network elements.

\subsection{Results from SuFI data}



\begin{figure*}
   \centering
   \includegraphics*[width=0.9\textwidth]{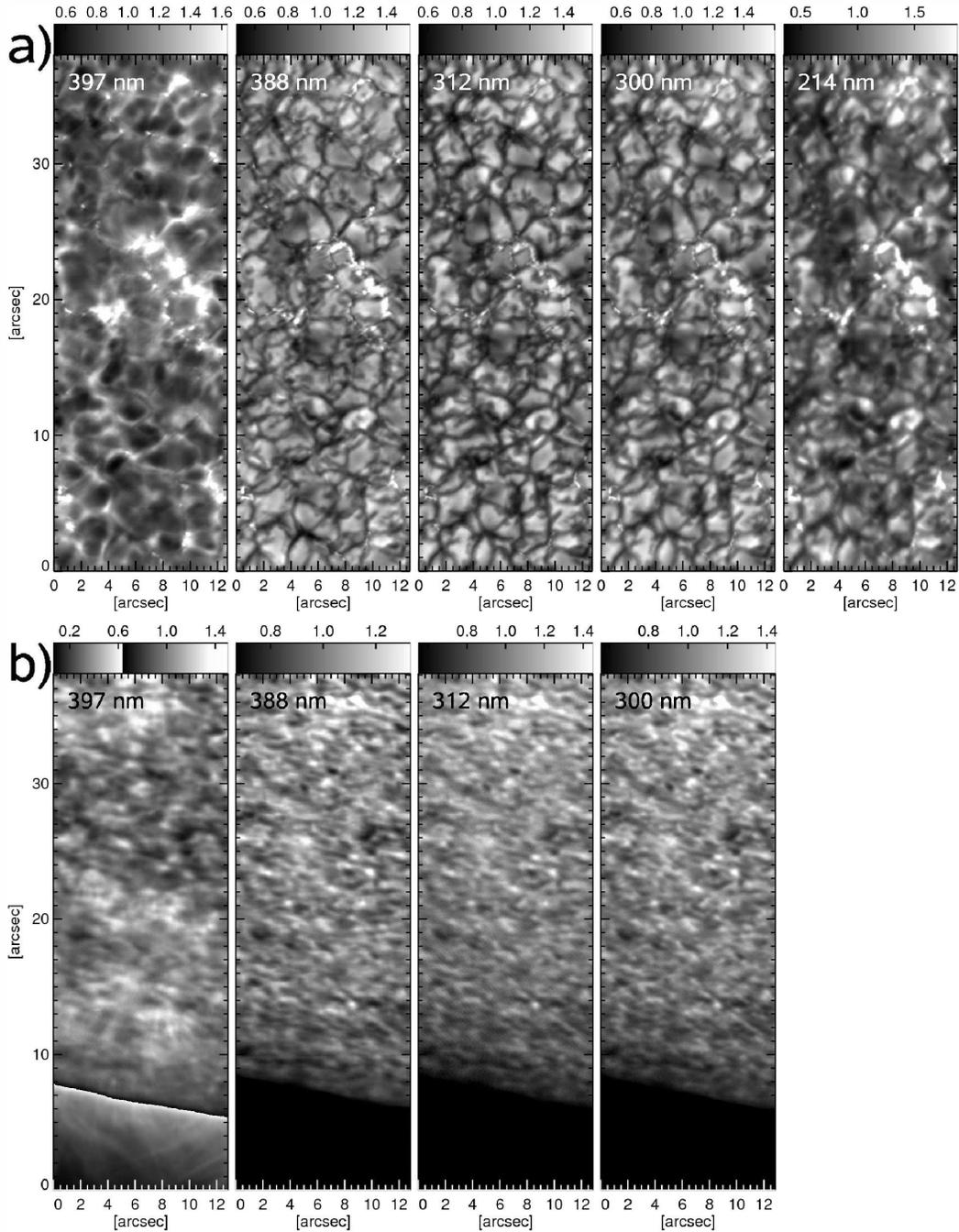}
   \caption{
   a) Images of a patch of quiet Sun near disk centre recorded by the SuFI
   instrument in wavelength bands centred on 397~nm, 388~nm, 312~nm, 300~nm and
    214~nm (from left to right). The grey scale has been individually set to
   cover 3 times the RMS range of each image.
   b) Same as panel a, but right at the solar limb. The
   Ca~H image (397~nm) is plotted with an enhanced brightness scale for the
   off-limb parts in order to reveal spicules. Owing to the low light
   level, no 214~nm data are available at this position.}
   \label{FigSufi}
\end{figure*}

Images of the quiet Sun at disk centre in all 5 SuFI wavelengths are
shown in Fig.~\ref{FigSufi}a, whose grey scale is saturated at
$\left\langle I\right\rangle \pm 3\sigma$ for each wavelength
in order to allow a better intercomparison of the granulation, at the cost of
overexposing the bright points (this figure shows the same solar scene as Fig.~13 of
Gandorfer et al. 2010, but with a different brightness scaling). The brightness
scale (see the gray-scale bars above the individual frames) already indicates
the large rms contrasts of the imaged granulation (see Hirzberger et al.,
2010a, for a quantitative analysis). The images at
300~nm, 312~nm and 388~nm display rather similar granulation patterns and prominent
bright points (BPs), especially in the network feature near the centre of the frame.
Even in the 214~nm image, granulation is well identified, although intergranular
lanes are no longer that clearly visible. It cannot be completely
ruled out, however, that this is due to a larger spatial smearing.
In addition, the granules display more substructure in the 214~nm images.
This is in contrast to the statement made by Hers\'e\ (1979) that ``at $\lambda 200$ nm
we find only bright grains of a mean area of 4 arcsec$^2$ and a mean contrast of
47\%''. Hers\'e\ identified them with facular grains. The reason for the difference
may partly lie in strong underexposure of the granulation (see his Figs.~4 and
5). There is a hint of granule-like structures in the lower right part of his Fig.~7
(only visible in the printed paper).

Bright points are particularly prominent at 214~nm (their contrasts are highest
at this wavelength, see Riethm\"uller et al., 2010), but somewhat more diffuse.
This may be due to image jitter accumulated during the 30~s
exposure and the greater height of formation of this wavelength, combined
with the expansion of magnetic features with height.


Ca~{\sc ii}~H~397 nm displays a variety of features including reversed
granulation, oscillations and waves, and magnetic bright points. The leftmost
frame of Fig.~1a reveals that although many of the intergranular lanes are indeed
bright, some appear as dark stripes with bright lanes on either side of them.
Examples can be found extending between the coordinates $(4, 1)$ and $(5, 3)$, from
$(4, 4)$ to $(3, 7)$, or from $(5, 8)$ to $(8, 7)$. Time series of Ca~{\sc ii}~H images
(see the movie M1 in the supplementary electronic material) reveals that these dark
Ca~{\sc ii}~H lanes are rather short lived. Possibly
they are a result of the interaction of the reversed granulation with waves.



At an intermediate heliocentric angle of $\mu=\cos\theta$ of roughly 0.72 (not shown)
the granulation appears very similar at 300, 312 and 388~nm. At 214~nm,
however, structures intermediate between reversed and normal granulation appear,
with bright lanes (on the centreward side) and darker granule bodies.

One interpretation of the different behaviour displayed by 214~nm data
is that the radiation in this band is formed (on average) at a greater height.
A rough estimate, made including line opacities, can be obtained from Fig.~1 of
Vernazza et al. (1976). It shows that around 300~nm the radiation is formed
roughly 50~km above $\tau_{\rm 500 nm}=1$, while at 214~nm it is formed around
250~km above $\tau_{\rm 500 nm}=1$. The true value at 214~nm is rather tricky
to determine owing to the broad filter profile, the presence of the
ionization edge of Aluminium within the filter range and the great density of
spectral lines at these
wavelengths. We speculate that 214~nm wavelength radiation is formed
sufficiently high in the atmosphere that first signs of reversed granulation
become visible at disk centre (e.g. filling in of intergranular lanes in Fig.~1a).
As we move towards the limb the reversed granulation becomes more apparent.

Reversed granulation is seen clearly in Ca~{\sc ii}~H, but many of the
finer-scale features are lost. The intermediate nature of 214~nm is illustrated
by the fact that the best correspondence of 397~nm is found
with 214~nm (coherence values are a factor of 1.5 higher than those
between, e.g., 300~nm and 397~nm).


Images at 4 SuFI wavelengths at the solar limb, in the general vicinity
of the south solar pole are shown in Fig.~1b.
Obviously the structures at the south limb, although reminiscent of granules,
are considerably smaller than those further away from the limb, in qualitative
agreement with earlier results (S{\'a}nchez Cuberes et al. 2003). This
smaller size could be related to greater prominence of granular sub-structure
at the limb. In addition, polar faculae are visible, e.g. around $y = 13$, 20
and 36. Fibrils can be seen emanating from the faculae at $y\approx 13$
and spicules are clearly present off the solar limb (they became visible after
the off-limb brightness was enhanced). Some of the spicule-like
structures appear strongly inclined to the vertical and even slightly bent,
so that they may actually be parts of quiet-Sun loops.

\subsection{Results from IMaX data}

\begin{figure*}
   \centering
   \includegraphics*[width=\textwidth]{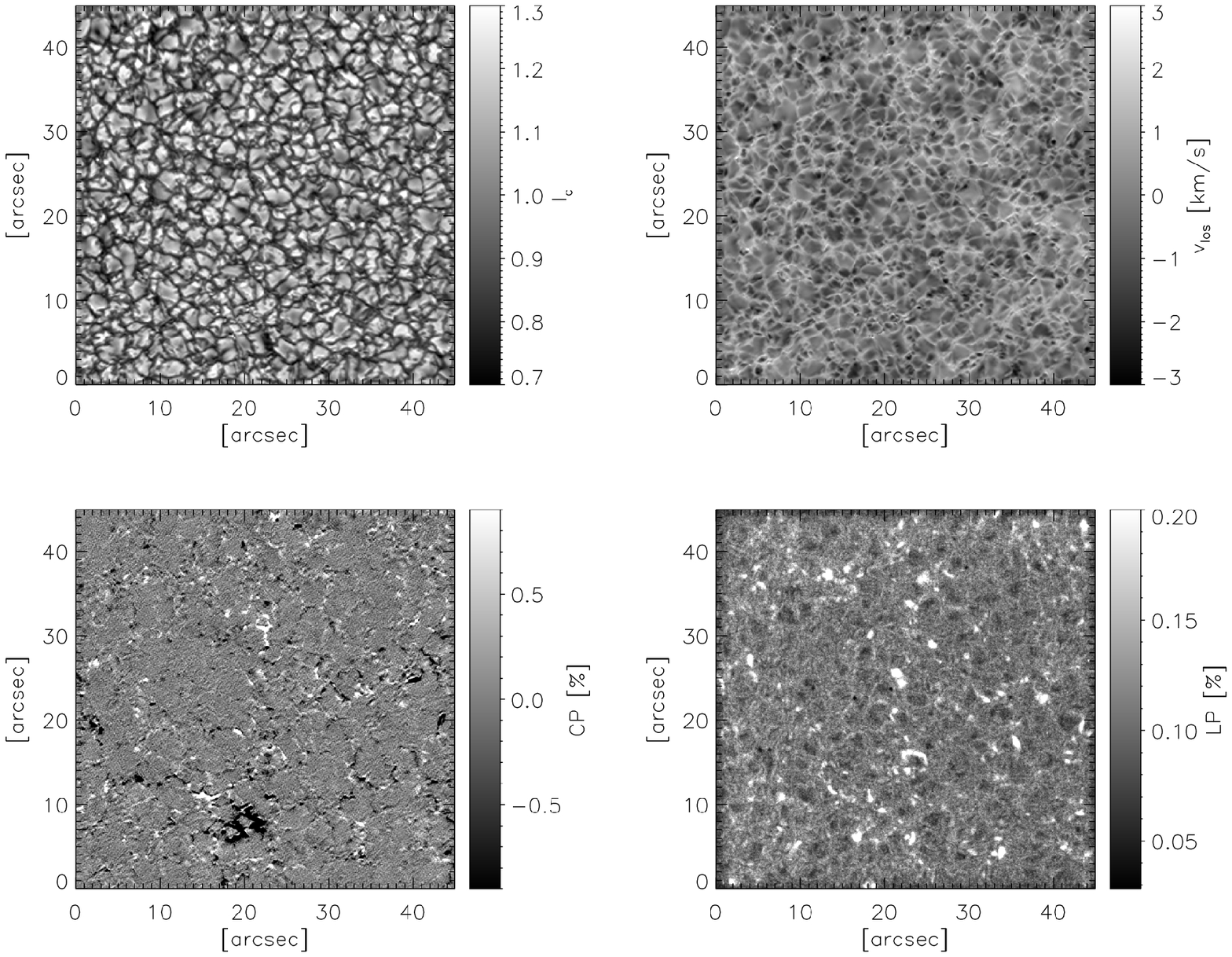}
   \caption{IMaX data. Clockwise from upper left: continuum intensity at 5250.4~\AA,
   line-of-sight velocity, net linear polarization $L_s$ (see text for
   an exact definition), line-averaged Stokes $V$, $V_s$ (see main text for a
   definition), obtained from Fe I 5250.2~\AA. All images are based on
   reconstructed data except for the linear polarization image.
     }
   \label{FigImax}
\end{figure*}

Figure~\ref{FigImax} shows a snapshot of IMaX data products; clockwise from upper left:
continuum intensity, LOS velocity, total net linear polarization averaged over
the line, $L_s$, and the similarly averaged Stokes $V$, $V_s$ (see below).
All quantities are based on reconstructed data, with the exception of $L_s$,
for which the unreconstructed data are shown (reconstruction increases the
noise, so that a number of significant $L_s$ patches in the unreconstructed
image are no longer sufficiently above the noise in the reconstructed data).
$V_s$ and $L_s$ are defined as:
 $$V_s={1\over 4\left\langle I_c\right\rangle}\sum^4_{i=1}a_iV_i\hbox{\rm \ \
 and\ \ }
 L_s={1\over 4\left\langle I_c\right\rangle}\sum^4_{i=1}\sqrt{Q^2_i+U^2_i} .$$
 Here $i$ runs over the 4 wavelength points inside the spectral line,
 $a_i$ takes on values of $1, 1, -1, -1$ for $i=1$ to 4 (see Mart\'{\i}nez
 Pillet et al., 2010) and $\left\langle I_c\right\rangle$ is the continuum
 intensity averaged over the FOV.

The RMS contrast of
quiet-Sun granulation obtained from IMaX continuum data is around 13.5\%, which
is a mark of the outstanding quality of the IMaX images. Due to the longer
wavelength at which IMaX observes, the contrast is lower than in the SuFI images. In
particular, the bright points are less clearly visible, having a contrast
comparable to that of granulation
(see Riethm\"uller et al., 2010). The velocity images prominently display a
network of sharp intergranular downflow lanes (appearing bright in the figure).
Upflow velocities are often largest in small patches, corresponding to small,
often bright granules or parts of larger granules, although there is no
one-to-one relationship between upflow speed and brightness (e.g. Hirzberger
2001). In addition, a larger-scale velocity pattern composed of 5--10$''$ patches
superposed on the granulation reflects the $p$-mode oscillations (see Roth
et al., 2010, for a detailed analysis of the $p$-modes, and Bello Gonz\'alez
et al., 2010, for a determination of the vertical energy flux transported by
high frequency waves).
These enhance the granular down- and upflows according to the oscillation
phase, as can be seen from the time series displayed in the movie M2
(in the additional electronic material).

The largest patch of strong Stokes~$V$ near the bottom centre of the frame is a
network element. Other network patches are also present in the frame, but are
less prominent, e.g. at (23,~31) in Fig.~2. Most of the Stokes~$V$ signal, however, is
due to internetwork fields. These are clearly composed of rather localized
(point-like or crinkled line-like) mixed-polarity magnetic patches. The sizes of
internetwork magnetic features deduced from Stokes~$V$ is typically below
1$''$, with many having sizes lying close to the spatial resolution of
roughly 0.15$''$. The internetwork fields clearly outline cell-like
structures with "mesogranular" scale sizes (cf. de Wijn et al. 2005; 2008)
that range from 2-3$''$ in regions with strong
internetwork flux to 5-15$''$ in regions with particularly low flux. There
are hints of smaller cells there, which, however, cannot be confirmed
without an in-depth analysis.
At many locations both magnetic polarities are located in close proximity to
each other.

$L_s$ displays a meso- to supergranular scale pattern.
The larger apparent scale of this pattern (compared to that displayed by
Stokes $V$) may result from the fact
that, on average, $L_s$ features have a lower S/N ratio than Stokes-$V$ patches,
so that we are probably missing more features of the former. The spatial
distribution is also different, with the most
prominent Stokes-$V$ elements being absent in the linear polarization
signal. The opposite is generally not the case. The more prominent patches of
linear polarization are usually associated with (weaker) patches of Stokes
$V$ (see Danilovic et al. 2010a).

Stokes-$V$ movies, such as that displayed in M2, reveal how dynamic
the quiet Sun magnetic field is, with the weaker magnetic features, i.e.,
those in the internetwork, being particularly dynamic. Constant appearance
and disappearance of patches of Stokes $V$ is observed along the edges of the
``mesogranular scale'' internetwork cells. As pointed out by de Wijn et al.
(2008), weaker features often disappear and
reappear close by. This could be an effect of features dropping
below the noise level by weakening and appearing again as they get more
concentrated. However, we also expect the emergence and submergence
of magnetic flux to take place on timescales of minutes. This is
supported by the analysis of $L_s$. Strong patches of linear polarization
are  found to be rather short-lived and are often associated with bipolar
magnetic features suggestive of small-scale loops (Danilovic et al. 2010a).
Some of these are also associated with supersonic velocities, presumably in
the form of upflows (Borrero et al. 2010).
There are also locations at which fresh flux emerges in a complex patch of
mixed polarities (e.g. Zhang et al. 1998), similar to the simulations of
Cheung et al. (2008),  but on a small scale. An example in our data is given
at around (36, 18) around the middle of the time series.

The {\sc Sunrise}/IMaX  Stokes-$V$ movie displays  qualitative
similarities to movies of the vertical magnetic field in turbulent dynamo
simulations (V\"ogler and Sch\"ussler 2007), although both the spatial and the
temporal scales are quite different. However, both display vortical motions of
weak magnetic field patches (see Bonet et al. 2010 for a study of
the vortices in these data; cf. Steiner et al. 2010). In the simulations, the mixed-polarity magnetic
fields are distributed between granules. There the individual magnetic features live less
than the granule lifetime. In the observations both size scales and lifetimes are
significantly larger/longer. Quiet-Sun magnetic fields in both, the turbulent dynamo
simulations and Hinode observations, display a
self-similar (fractal) spatial distribution (Pietarila Graham et al. 2009).
This supports the idea that structure and evolution over different scales
(of our observations and the MHD simulations) are similar.

The unreconstructed Stokes $V$ map has a noise level of 3.0 G, with a
conversion from Stokes~$V$ to $B$ following Mart\'{\i}nez Pillet et al. (2010).
Similarly, the reconstructed map has 9.8 G. The average $V$ above
$2\sigma$ in the reconstructed and non-reconstructed maps (i.e. averaged
over all pixels, but with the signal set to zero in all pixels with
$V< 2\sigma$), converted into LOS field component, $B_z$, gives
$\langle |B_{z,\rm rec}|\rangle =7.31$ G above $2 \sigma_{\rm rec}= 2\cdot{} 9.8$
G and $\langle |B_{z,\rm nonrec}|\rangle = 4.01$ G for $2 \sigma_{\rm nonrec}=
2\cdot{} 3.0$~G. The index ``rec" implies reconstructed data, ``nonrec"
non-reconstructed. Thus, in spite of the more than 3 times higher noise, the
reconstructed
image displays almost twice the amount of magnetic flux. If we set the same
limit for both maps, namely $2 \sigma_{\rm rec}$, then
$\langle |B_{z,\rm nonrec}|\rangle$ is reduced to 1.96 G. This means that
3.7 times more magnetic flux is found if the effective spatial resolution is
increased by roughly a factor of 2. This difference persists also if the
threshold is set higher, so that it cannot be due to contributions from pixels
containing noise $ > 2 \sigma$.

This strong increase in magnetic flux suggests that the magnetic polarities
are mixed at very small scales approaching the resolution limit of the present
data, so that even moderate additional smearing leads to
a substantial decrease in the amount of visible flux.
This dependence on resolution is much stronger than found for
the quiet Sun at lower spatial resolution by Krivova and Solanki (2004),
and seemingly also than found by Pietarila-Graham et al. (2009). Quantitative
comparisons are difficult, however, since the
reconstruction does not, strictly speaking, change the spatial resolution,
but rather significantly enhances the MTF at frequencies below the cutoff.

The increase in magnetic field strength also implies a considerably higher
magnetic energy in the chromosphere (Wiegelmann et al. 2010).

\subsection{Combination of SuFI and IMaX data}

\begin{figure*}
   \centering
   \includegraphics*[width=0.9\textwidth]{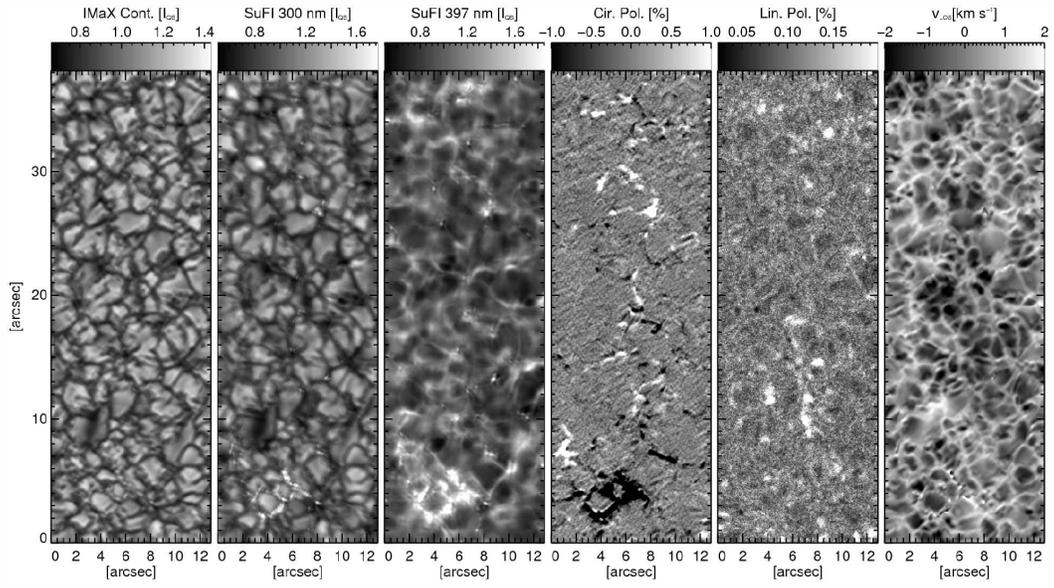}
   \caption{Cotemporal SuFI and IMaX data showing the SuFI field of view.
   From left to right: continuum intensity at 525.04~nm (IMaX data),
   intensity in the 300~nm band (SuFI data), intensity in the Ca~{\sc ii}~H line
   core (SuFI data), net circular polarization deduced from Fe I 525.02~nm
   (IMaX data), total net linear polarization ($L_s$; IMaX data), line-of-sight
   velocity (IMaX). All images have been reconstructed except for
   the linear-polarization image.}
   \label{FigSufiImax}
\end{figure*}

Figure~\ref{FigSufiImax} displays a set of cotemporal and cospatial SuFI and IMaX data.
Clearly, the prominent, small-scale photospheric bright points in the
photospheric SuFI image (2nd frame from left) are all associated with
brightenings in Ca~{\sc ii}~H and with Stokes-$V$ signals, although
in some cases these signals are not particularly strong (cf. Riethm\"uller et al.
2010). Note that in the internetwork these bright points are still very sharp
also in Ca~{\sc ii}~H, which distinguishes them from more wave-like features,
which are in general not quite so concentrated. In the 525~nm
continuum, however, they are barely detectable. After identifying them in the
300~nm image they can, however, be made out as points with a similar
brightness as the surrounding granules.

The relationship between Ca~{\sc ii}~H brightness and magnetic field is more complex.
While some of the magnetic features are associated with Ca~{\sc ii}~H brightenings,
e.g. in the network feature near the bottom of the frame, or other magnetic
features related to internetwork bright points, there are numerous magnetic
features in the internetwork, e.g. at (3.5,~8.5) or between (7.5,~8.5) and
(7.5,~11), which are not associated with Ca~{\sc ii}~H brightenings. Such Ca-dark
magnetic features  often display mixed polarity and are closely related to patches of $L_s$
(one striking case is discussed by Danilovic et al. 2010b). Conversely, there
are numerous Ca~{\sc ii}~H brightenings
which do not seem to have any magnetic counterpart in the photosphere.
Examples are at (12,~14) and at (10,~22). Although some of them can appear
relatively concentrated, on the whole they are clearly more diffuse than the
brightenings due to magnetic features. These may correspond to H$_{2V}$ bright
points (Rutten and Uitenbroek 1991), although without further spectral
information we cannot be sure.

\section{Conclusion}

The {\sc Sunrise} observatory has provided high-quality,
high-resolution images,
Dopplergrams and vector magnetograms at different positions on the solar
disk. The extremely low solar activity level at that time means that these data
mainly enable new insights into the quiet Sun. Here we provided a qualitative
description of these data and some of the features visible in them. This, together
with the more quantitative analyses described in the following papers in this
special issue, has already led to new insights into the magnetism, convection
and oscillations and waves in the quiet Sun. Given the richness and
quality of the data and the fact that so far only a very small fraction of
them have been analyzed, we expect many more results to follow. A flight of
{\sc Sunrise} during a period of higher solar activity is greatly to be
welcomed.

\begin{acknowledgements}
We thank R.~Rutten for helpful discussions on the structure of the
chromosphere.
The German contribution to {\sc Sunrise} is funded
by the Bundesministerium f\"ur Wirtschaft und Technologie through
Deutsches Zentrum f\"ur Luft- und Raumfahrt e.V. (DLR), Grant No.
50 OU 0401, and by the Innovationsfond of the President of the
Max Planck Society (MPG). The Spanish contribution has been
funded by the Spanish MICINN under projects ESP2006-13030-C06
and AYA2009-14105-C06 (including European FEDER funds). The
HAO contribution was partly funded through NASA grant number
NNX08AH38G.
This work has been partially supported by the WCU grant No. R31-10016 funded
by the Korean Ministry of Education, Science \& Technology.
\end{acknowledgements}

\section{Electronic supplementary material}

Movie M1: Timeseries of Ca~{\sc ii} H images recorded by SuFI on Sunrise. The
movie runs for 19 min solar time (see counter at the top). The brightness
scale (grey scale at the right) is given relative to the brightness averaged
over the time series.

\vskip 0.5cm

Movie M2: Continuum intensity, line-of-sight velocity, linear polarization
$L_s$, and circular polarization $V_s$ deduced from IMaX measurements in the
Fe I 525.02 nm line (clockwise from top left). See
the main text (Sect. 4.2) for the definitions of $L_s$ and $V_s$.
The movie shows a region around $48\times 48$ arcsec in size. The edges have
been slightly cut-off, since they are reduced by the reconstruction (this can
partly still be seen in the various plotted quantities).

\end{document}